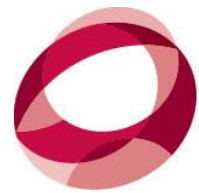

# Research Opportunities and Visions for Smart and Pervasive Health

Elizabeth Mynatt
mynatt@cc.gatech.edu
Georgia Tech

Gregory D. Hager
hager@cs.jhu.edu
Johns Hopkins University

Santosh Kumar
skumar4@memphis.edu
University of Memphis

Ming Lin
lin@cs.unc.edu
University of North Carolina
Chapel Hill

Shwetak Patel
shwetak@cs.washington.edu
University of Washington

Jack Stankovic
stankovic@cs.virginia.edu
University of Virginia

Helen Wright
hwright@cra.org
Computing Community Consortium
June 15, 2017[1]

Improving the health of the nation's population and increasing the capabilities of the US healthcare system to support diagnosis, treatment, and prevention of disease is a critical national and societal priority. In the past decade, tremendous advances in expanding computing capabilities—sensors, data analytics, networks, advanced imaging, and cyber-physical systems—have, and will continue to, enhance healthcare and health research, with resulting improvements in health and wellness. However, the cost and complexity of healthcare continues to rise alongside the impact of poor health on productivity and quality of life. What is lacking are transformative capabilities that address significant health and healthcare trends: the growing demands and costs of chronic disease, the greater responsibility placed on patients and informal caregivers, and the increasing complexity of health challenges in the US, including mental health, that are deeply rooted in a person's social and environmental context.

Emerging paradigms in smart and pervasive health recognize that health, and a growing percentage of healthcare, takes place outside the traditional walls of clinical care and is governed by the day-to-day experiences that make up the fabric of daily life. Diabetes costs the nation $245 billion each year, yet prevention and management of diabetes is inextricably tied to a multitude of daily decisions and stresses.[2] Likewise heart disease generates an even larger

---

[1] This paper is an outgrowth of the workshop Discovery and Innovation in Smart and Pervasive Health sponsored by the Computing Community Consortium (CCC) on December 5-6th, 2016 in Washington, DC. This workshop brought together leading researchers and policymakers to discuss the successes of Smart and Pervasive Health research activities, the evolution of relevant computing capabilities and the application of these technical innovations to health and wellness goals. This report highlights these paradigms and concludes with specific recommendations for the successful future implementation of an impactful smart and pervasive health research agenda.

Thanks to Kevin Fu, Nina Mishra, Katie Siek, Donna Spruijt-Metz and Greg Welch for their helpful comments and editorial advice.

Contact: Ann Drobnis, Director, Computing Community Consortium (202-266-2936, adrobnis@cra.org). For the most recent version of this paper, as well as related reports visit: http://cra.org/ccc/resources/workshop-reports/

[2] (2015, June 22). The Cost of Diabetes.... Retrieved April 26, 2017, from http://www.diabetes.org/advocacy/news-events/cost-of-diabetes.html

economic impact of $317 billion annually.[3] Diagnosis and treatment of chronic conditions such as asthma and epilepsy are stymied by huge gaps in environmental information and incomplete knowledge of notable events such as asthma attacks and seizures. Moreover, any improvements tied to behavior changes such as improved nutrition, exercise and sleep and the cessation of behaviors such as smoking, require personalized interventions tailored and responsive to a person's environmental, social and psychological context. Finally, the growing crisis in mental health also requires a more accurate understanding of environmental and social influences and the ability to create targeted and timely personalized interventions.

Moving the needle on healthcare costs and outcomes requires embracing this new frontier of data, analytics, mobile and in-situ interventions, and integrated human-centered systems. Improved clinical care and discovery require developing new scientific foundations that integrate new data available from the "exposome" that surrounds and influences human behavior and health.[4] Public health, rooted in the need for accurate interpretation of available data and effective communication, also has much to gain from the growing availability and robustness of sensors, mobile health capabilities, data analytics, and human-system modeling described in this paper. Individuals desiring to take care of their own health, and the health of family members, stand to benefit from evidence-based home and mobile applications that are safe and secure.

Realizing the potential of smart and pervasive health and healthcare approaches requires dedicated and sustained investments in basic computing research in concert with mission-focused research that narrows the gap between technology and scientific computing advances and positive health outcomes. In this paper, we describe three critical areas for basic computing research: heterogeneous sensing and data analytics; integrated cyber-human-physical systems; and comprehensive physical and digital security capabilities. Taken together innovations in these computing capabilities have the potential to transform US health and healthcare while also advancing fundamental computing research and contributions to other domains.

**1.0    Computing Advances: Holistic Sensing, Advanced Analytics and Comprehensive Decision Support**
Always available health sensing and behavioral monitoring offers the transformational possibility of advancing health and wellness that is bound by neither space nor time. This new transformation in health technologies enables observations of dynamic changes in each individual's health state, as well as key physical, biological, behavioral, social, and environmental factors that contribute to health and disease risk, anytime and anywhere that have not been possible in the past.

Although there has been a significant progress in the use of health sensors for remote care for specific diseases (e.g. heart rate monitors), **future strides will come from "holistic" sensing approaches** that **integrate wearable, environmental, behavioral, and social network information** and create new computational models that can dynamically recognize the influence of environmental factors on human physiology and behavior. Much of the current work in health sensing has focused on point solutions for specific disease conditions or states, and has largely been informed by conventional diagnostic techniques that do not take into account environmental and social information. For example, with the explosive growth of social media data, it is now possible to obtain temporally varying measurements of interpersonal exchanges, social network structures, and social capital, which can be combined with complementary physiological and psychological sensor measurements of an individual. These dynamic measurements can also reveal network and community effects in health outcomes — quantification of which has been challenging so far due to the paucity of adequate social and community-level data. Collectively this holistic picture of a person's health and health-related behaviors complements today's genomic and clinical data while honing in on the social and environmental factors that primarily contribute to disease onset and progression and that pose barriers to good health.

---

[3] (2017). Costs & Consequences.... Retrieved April 26, 2017, from https://millionhearts.hhs.gov/learn-prevent/cost-consequences.html

[4] (2014, April 21). Exposome and Exposomics…. Retrieved May 1, 2017, from https://www.cdc.gov/niosh/topics/exposome/

These approaches also aim to capture the *context* surrounding human behavior to effectively guide healthcare delivery. For example, the noise and bustle of an ICU should affect the content of ICU information displays, just as the availability of healthy food and measures of social influence should affect diabetic care and nutritional coaching. In another example, given the range of potential trigger events for PTSD, effective treatment plans for an individual would benefit from historical and real-time awareness of a person's social and environmental context. This combination of approaches, integrating information about a person's physiological and psychological state with their environmental and social context can enable the creation of targeted and personalized care to address numerous health challenges.

These approaches require addressing several challenging issues in analyzing data, designing decision-support systems, and developing strategic mechanisms for reaching informed responses from holistic sensing:
- How can we measure key environmental variables, such as exposure to cues and triggers for adverse health-related behaviors, and infer a comprehensive characterization of individual behavior?
- How can we model the effect of environmental and social factors on behavior regulation to support behavior change and behavior maintenance?
- How do we capture and effectively model the context of and around an individual to guide "hows" of health promotion and health care delivery: e.g. how to best deliver information, how to best deliver interventions where and when they are needed, how to support caregivers, or how to best communicate with a patient?
- How can we support decision-making where prioritization of key factors, for example emotional vs. physical well-being or invasive vs. non-invasive treatment, is critical to effective health promotion and clinical care?
- How can we improve robustness and reliability of these systems operating "in the wild" outside of the traditional confines of healthcare environments?

In turn this application-driven research exposes basic computing research challenges that are relevant across many domains including:
- Combining large-scale population "big data" with temporally dense as well as sparse individual data, controlling for biases in each;
- Visualizing context-aware mixed-modality data analytics and high-level abstraction and summarization of large-scale, multi-modality data with uncertainties;
- Designing time-series pattern mining methods to extract events of interest from multi-modality data stream with varying temporal granularity, spatial irregularity, varying reliability and validity, and data incompleteness; and
- Developing provenance systems that capture both metadata and annotations of the entire data processing stage to facilitate both interpretability and comparative analysis;

In summary, these fundamental computing research challenges, when explored in the context of healthcare delivery, health and wellness, have the tremendous potential to address both long-standing national priorities in health and catalyze significant research advances in computing.

**2.0     Computing Advances: Human-Centric System Approaches for Smart and Pervasive Health**
Holistic data enables a spectrum of system capabilities that integrate data insights and support human action. **Fundamental progress will come from the creation of systems that actively incorporate input from participants, patients, healthcare providers and caregivers; that provide pathways for action, ranging from** **automation to decision support; and that emphasize human engagement** as a whole. These systems must grapple with a host of challenges including balancing human direction and autonomy, integrating real time input and control, creating scalable approaches that manage overlapping layers of input and action, and creating mechanisms for learning. Here we outline key **system** approaches that requires fundamental advances in computing research to drive transformative health and healthcare solutions.

**Mixed Initiative Systems and Closed Loop Systems:** Mixed initiative systems balance automation and independent human action including patients, physicians and caregivers to achieve a desired goal. For example, consider a wearable system that helps monitor activities and glucose for patients with diabetes. This system could provide a clinician a weekly summary of a patient's daily activity and note unusual spikes and dips in glucose; it might offer supportive hints and reminders throughout the day to the patient; and send requests for help or feedback to family members as necessary. Mixed initiative systems must also create flexible and time-varying divisions of responsibilities. Decisions that a patient might can make under typical conditions might not be possible under periods of stress, or during episodes of depression or days when dementia symptoms are acute. While traditional mixed-autonomy systems rely on a single, engineered structure for division of task and responsibility, next-generation systems must have the capability to safely support different models of responsibility and risk management for individualized care communities and across changes in the capabilities of individuals. Mixed-initiative systems must also have some underlying model of possible human action. Anticipating the range of "random" human action is a challenge, especially when the goal support fully automated systems.

Systems that innovatively aim to "close the loop" – using data, analytics, and interaction with patients and providers to adapt to dynamic situations – help ensure responsive action as health conditions and needs change. Continuing the above example, a problematic glucose reading could automatically trigger a phone call to a caregiver or real-time adjustment to a wearable insulin pump. Holistic sensing and analytics, as described previously, provide the needed information to these systems; indicating for example that a glucose reading is acceptable based on recent food intake and past history, or if a default caregiver is unavailable. However, adding real-time elements introduces the additional complexity of trading automation/autonomy with oversight, filtering, and feedback through people.

Furthermore, creating mixed-initiative and closed-loops systems typically involves choosing an objective to optimize, but in smart health, which objectives do we chose, and how do we measure them? For example, rehabilitation may be accelerated by "pushing the system" to the limit, but at the risk of patient comfort; coaching may lose effectiveness if it does not respect patient education, experience, bias, and values, things that are hard to measure and predict; likewise clinical decision-support may not be effective if it cannot take provider education and experience into account.

In addition, many of these systems are distributed – e.g. an ICU is really a complex cyber-physical system-of-systems with multiple controllers (decision makers) that either decide independently on different aspects of the overall system or collaboratively for the entire system, which raises the overall complexity level. Hence there are additional needs for approaches that enable the coordination of a system of systems working together, yet likely retain independent goals, controls and actions.

**Multi-Tiered Sensing, Modeling and Control Systems:** Multi-tiered Sensing, Modeling and Control (SMC) system approaches create the foundation for mixed initiative systems that flexibly balance automation and human action and closed loop systems that integrate real-time data to optimize for specific health outcomes. Healthcare data, computed inferences and associated models have a natural multi-tier structure, with varying levels of sampling rate, time-scales, abstraction and specificity. Therefore, it is natural to design "smart health" systems that mirror this multi-tier structure. A multi-tiered SMC framework opens opportunities for:

- Adaptively selecting different models at different time scales and abstraction to flexibly integrate multiple sources of data, from real time sensing to public health data.
- Enabling transitions from population models of behavior, seeded with data accrued at a population level, often static and based on sparse measurement, to individual, dynamic, learning models of behavior that incorporate temporally dense, contextualized data accrued the individual level.
- Enabling transitions from interventions that operate on population-based data to those that gradually incorporate more individual data and shift to personalized interventions based on dynamic, learning models of the person.
- Balancing tradeoffs in sensing, model prediction, data sharing, and privacy needs.
- Engineering systems to optimize resources across multiple data gathering, modeling, and decision-making processes.
- Dynamically and iteratively gathering information as needed, such as incrementally gathering contextual information as a person's activity changes throughout the day.

The realization of mixed initiative, multi-tiered and closed-loop systems for healthcare, including innovations in fundamental sciences and new technology, introduces both promises and challenges. For example, some research issues include real-time data processing, modular and reconfigurable system design, interactive human-in-the-loop feedback-control strategies, formal methods for safety specification, reliability analysis, fault-tolerant dependability and robustness techniques for coping with partially known environments with high level of uncertainty. With cyber physical systems technologies, closing the loop via interventions poses new problems. Interventions will span the spectrum from single cell control, to on-body control devices, to in-home and in-city environments, and even for public health concerns such as pandemics.

Ideally these systems learn over time, but there is not a single loop in most cases, but rather multiple loops at different levels of reaction time and generality. For example, the human body has its own biology based loops to detect and react to disease. New work aims to use control loops to affect individual cells. Other "inner loops" are more traditional real-time control loops (e.g., technology assessing and adjusting prosthetics in real-time based on use), "mid-tier loops" being more human-controlled loops (e.g., adjustment of prosthetics settings by physician) and much slower or less frequent, and "outer loops" producing generalized knowledge (e.g. likely outcomes for different types of prosthetics). Advice, alarm, and recommendations can be considered other forms of actuation, hence are aim to close the loop.

It is important to note that interventions can be "textual," e.g., in the form of alarms and reminders. Hence, advances in natural language processing is required to understand the semantics of textual information available to or sent to patients, including personalizing instructions and filtering and correcting any unsafe actions. This capability requires processing in the control loop to understand the semantics of text, filter the text for the right information for the individual, and to avoid conflicting and harmful advice. Moreover interventions can utilize haptic, visual and other physical cues requiring increased design knowledge for the effective implementations of these systems.

In summary, we call attention to these fundamental computing research challenges that when explored in the context of health have the potential to address long-standing national priorities in health and healthcare. Additional challenges that span these approaches include:
- Modeling complex dynamical systems with uncertainties from multiple sources to allow inference making in the presence of temporal and spatial imprecision;
- Understanding how are inferences and predictions created, how is uncertainty represented, and how can humans understand the rationale and limits of these systems? For example, a neural net may produce outstanding predictions, but without clear justification or "error bounds"; an explicit model may be more explainable and, although it has reduced prediction performance, may be more "convincing" to a provider or patient.
- Deriving predictive models that efficiently process multi-modality data, incorporate multi-dimensional data analytics, and perform multi-variable statistical inferences from "digital biomarkers" to facilitate decision making;
- Integrating analytical techniques (machine learning, deep learning, data mining, Bayesian reasoning), and dynamical system and control system theories with cyber physical systems for real-time control.

**3.0      Integrating Advanced Computing Capabilities into Healthcare Delivery**

We briefly present two healthcare delivery challenges that illustrate the significant potential benefit from advanced holistic data analytics and multi-tiered, mixed-initiative, closed loop systems.

**Distributed Coordinated Care Delivery:** Healthcare, particularly for the chronically ill, presents enormous challenges for coordinating care. A recent NAM report notes that a single care provider may be coordinating care with over 200 other clinicians.[5] Compound these interactions with nurses, rehabilitation specialists, and technicians, all operating in parallel to treat many patients through activity on thousands of tasks, and the complexity of care coordination explodes. The opportunity space for technological advances in this area is rich, with a particular focus on simplifying workflows, and amplifying human capabilities through computational means.

To make this concrete, imagine an overarching care coordination framework that is able to track patients, caregivers, and activities both within and outside the care environment. This framework is aware of standard workflows for patients and is also aware of patient inflow and worker capability. To create and field such a system, research is needed on a wide variety of topics, including:

- **Modeling Caregiver Capabilities and Human Limitations:** Consider the range of caregivers -- from trained professionals to family and friends. An intelligent coordination system should be able to model the types of care and support (from emotional to physical to medical) that each can provide, and provide an optimal assignment of healthcare workers to a given patient population. Options may be in-patient intensive care, out-patient rehabilitation, or health monitoring. These models should be able to be personalized, and thus require constant adaptation and learning from the caregivers they represent. They must also recognize human limitations to ensure the health and well-being of caregivers as well as patients.
- **Modeling Empirical Workflows:** As Lord Kelvin famously stated, that which cannot be measured cannot be improved. Most healthcare proceeds along standardized care pathways, but care diverges due to patient variability, caregiver choices, and local resource constraints, to name a few examples. Smart health systems will need to develop and adapt care pathways to the characteristics of patients, and caregivers, and be coupled to data analytics to model and assess the value of such pathways, as well as propose improvements for such pathways.
- **Dynamic Scheduling and Assignment:** Following from the above, how do we now dynamically and optimally assign role to humans, and transfer the necessary information between team elements to ensure smooth and safe transition? How can potential divergences be rapidly detected and team-members alerted before errors occur?
- **Effective Assistance:** There are myriad details related to care -- has a patient had their medication? Have they moved recently? Should their bed be moved? Have they had human contact in the past few hours? Are they progressing as expected, or should they be attended to? Most of these tasks that are easily monitored automatically and would greatly relieve points of stress or error from patient care. However, such assistants must understand care expectations, be able to sense and understand actions in and around a patient, and be engineered to provide useful and actionable information in easy-to-comprehend ways to the right person. Research in systems for distributed sensing and human-machine interaction will be needed to develop these ideas.

---

[5] (2013). Best Care at Lower Cost: The Path to Continuously Learning Health Care in America … Retrieved May 1, 2017, from https://www.nap.edu/catalog/13444/best-care-at-lower-cost-the-path-to-continuously-learning

**Just-In-Time Adaptive Interventions:** Just-in-time adaptive interventions (JITAI, pronounced "jedi") integrate mobile health applications with dynamic information about an individual's emotional, social, physical and contextual state to facilitate the adoption and maintenance of healthy behaviors, and to discourage and prevent negative health behaviors and outcomes. They are designed to provide the right type and amount of support, at the right time, in person's natural environment by adapting to the person's changing internal state and context. These interventions draw heavily from the research in modeling and the creation of human-in-the-loop, mixed initiative systems.

For example, most (93%) unaided smoking cessation attempts fail in first week and 95% of lapses lead to a relapse to smoking. Stress, craving, alcohol, and proximity to tobacco outlets are some major precipitants. The challenge is how to monitor a dynamic environmental context in real-time to trigger timely interventions that are matched to real-time and historical behavioral information. To create and field such a system, research is needed on a wide variety of topics, including:

- **Effective Interventions:** JITAI systems must constantly address the challenge of understanding what intervention (e.g. a reminder of a past health goal) could be effective at a particular moment in time (e.g. traveling for business) given a specific physiological and psychological context (e.g. stress, hunger, separation from social support). Interventions that are delivered too late (e.g. after ordering a large pizza), are too annoying or otherwise out of tune with the current context will be ineffective. The success of JITAI interventions hinges on understanding what generally works for an individual and then how to moderate or calibrate those interventions based on a specific context, as well as regularly 'tuning up' the model to be congruent with changes and development in the individual.
- **Balancing Tradeoffs of Action and Inaction:** Akin to understanding what interventions work well for an individual is also knowing how often to provide an intervention. Individuals can become annoyed and acclimatized with too frequent reminders, such as a buzzing pedometer to remind someone they have been sitting too long, or frequent offers of moments of mediation to combat stress. The consequences of an unhealthy action must be considered. Simply having one cigarette may be catastrophic for smoking cessation goals; resuming drug use while in recovery can be fatal. While the distinction between opioid use and not meeting a step count goal is obvious, tradeoffs surrounding stress, depression and mental health require more sophisticated reasoning techniques to inform the design of JITAI systems.
- **Incremental Context Gathering:** JITAI systems dynamically and iteratively gather information as needed, such as incrementally gathering contextual information as a person's activity changes throughout the day. Uncertainty, such as traveling to a new place, or surpassing threshold a key threshold, such as an overly stressful day, will drive the need to gather additional context to inform any interventions.
- **Training JITAI Systems:** The initial use of a JITAI system can often be critical to its success, but such a system, "out of the box", would lack needed training data to inform intervention choice and timing. What is needed are interactive surveys, quizzes and games that can be used to glean information about personal preferences, triggers and reactions to bootstrap JITAI interventions.
- **Interfaces to Care Networks**: Although the design of JITAI systems focuses on creating individualized support, they nevertheless can operate as part of a larger care network. Understanding how to disseminate information to this network and activating or delegating action to the network is an important need. Examples include alerting a healthcare provider to possible breakdowns in mental health, keeping family members apprised as a child with asthma transitions to college and more independent care, or rallying friends as part of a social competition or support for healthy behavior.

**4.0      Computing Advances: Security Challenges for Smart and Pervasive Computing**

Digital health technologies and computation are now permeating through all of healthcare – wireless nursing stations, insulin pumps, implantable medical devices, home blood pressure monitoring systems, health records, etc. To inform the design of these systems and to inform regulation, a comprehensive approach to cybersecurity in health needs to be considered. Many of the security and privacy vulnerabilities found in traditional computing technologies apply in the health setting, but they are amplified through the significant use of legacy systems, out of date operating systems, and many entry points for malicious activity. This range of entry points for attack is further complicated by the fact that personal health monitoring devices are constantly entering and leaving homes and clinics, thus moving away from isolated infrastructures and making classic technologies like firewalls and antivirus less effective.

Security and privacy in health is also about risk management and research is needed in how we balance the risk and benefits of these digital systems. Best practices in security and privacy also need to be re-considered for providers, patients, and developers given the workflows in healthcare. As the number and connectivity of such devices increases, the challenge of managing these collections of devices also becomes exponentially more difficult. Making a single device secure and safe is already a difficult problem. Safety issues are increasingly important for connected devices as many of these emerging systems are capable of physical control. Health sensing solutions are especially vulnerable as many of these systems reside near or on the body of a user. Past work has already shown the security vulnerabilities of implantable medical devices.[6] However, it is important to balance the clinical benefits of a device with the security risks such that the outcome is safer and more effective health and wellness rather than simply security as the end goal.

Data privacy and security tools for clinical research also need to be considered as the computing community's interest in the application of computing to health continues to grow. The increased volume and ease at which clinical and physiological data can be collected with smartphones and wearable sensors increases the risk of data breaches, accidental exposure, and violation of health privacy standards. Mechanisms for auditing digital health technologies need to emerge as regulatory standards start to be developed in this space. Users may not be aware if their health device is compromised, as well as, they may not be aware of where sensitive personal health information may be going.

**5.0      Supporting New Health and Healthcare Paradigms**

Advances in smart and pervasive health research create opportunities to support new paradigms in healthcare delivery and to address persistent healthcare disparities in the US. Specifically, these new holistic and human-centric approaches can provide the foundation for accountable and patient-centric care and for addressing systemic healthcare disparities based in inequities in environments, resources, education and access to healthcare. Some examples of new directions these technologies enable are the following:

**Value-Based Treatment Plans:** Currently, much of medicine is moving toward "outcomes-based" approaches to enhancing value – get the "best" outcome for the resources expended. However, this premise assumes there is a global and uniform definition of "value" – but value can be interpreted at many levels – patient value, family value, provider value, health system value, and economic value. For example, end of life is a focus point for value – a time of high expenditure with a perception of high value related to treatments (or lack thereof). What are lacking are tools that model "quality of life" alongside predictive models for healthcare treatments. Value-based treatment plans could also foreground the use of incentives in chronic disease management. For example, systems that coach patients with diabetes to sustain life goals such as outdoor activity and interaction with family members could prove much more effective than the more abstract task of managing glucose levels.

---

[6] (2016, Dec 27). Pacemakers CAN be hacked: US government issues cybersecurity warning over hackers programming bugs in medical devices... -NCBI. Retrieved May 9, 2017, from http://www.dailymail.co.uk/health/article-4068988/U-S-posts-rules-addressing-cyber-bugs-medical-devices.html

**Accountable and Patient-Centered Care:** Multi-tiered SMC systems can provide the foundation for accountable and patient-centric care by integrating complementary models for data collection, decision-making and collaborative, coordinated care. Likewise, systems that integrate more holistic information about patients' lives and information about networks of informal caregivers have the potential to shift current models of patient-centered from supporting physician collaboration to literally centering care around a patient's behavioral, social and environmental context.

**Informing Public Health**: Transformational approaches in smart and pervasive care have the potential to be the eyes and ears of a more informed public health system. Insights from pervasive monitoring can help inform research understanding the environmental role in conditions such as asthma, diabetes and heart disease, and help bolster public policy focused on health and wellness. Systems that center on human interaction and provide personalized interventions can catalyze effective approaches to sustain wellness and prevent disease. Finally, increased sophistication in behavioral monitoring and understanding the role of social influence can inform education, outreach and communication activities focused on improving the health of the US population.

**Decreasing Healthcare Disparities:** Different groups have been subject to, and suffer from inequities in, environments, resources, education, access to health care amongst other challenges, which can all adversely impact health outcomes such as quality of life and life expectancy.[7] These groups include, but are not limited to, different racial, ethnic and cultural groups, low socioeconomic status, low resource, women, LGBTQI, people who have physical and cognitive and sensory impairments, older adults, veterans, rural populations, etc. Smart and pervasive healthcare systems have the potential to reduce disparities by creating systems that are situated in specific socio-economic contexts, for example providing nutritional and exercise guidance in the context of local resources and economic constraints. Moreover, pervasive rates of smartphone use create a more equitable playing field for access to mobile sensing and healthcare expertise.

## 6.0  Cross Cutting Issues and Barriers to Success

Despite the promise and potential of smart and pervasive health capabilities, there are nonetheless several critical issues to address as possible barriers to success for the research community and for the long-term goal of providing meaningful health outcomes.

**Infrastructure Challenges: Platforms for Smart Health Research:** Sensing and actuation form the basis for smart and connected health research. More and more wellness and disease-specific studies are being conducted in homes and assisted living facilities. However, many of these research efforts build one-off platforms. This approach is very inefficient and subject to error and inconsistency. What is required is common infrastructure that is easily modifiable for different health studies. This common infrastructure should include, at least, sensing and actuation libraries with associated software processing, networking and communication capabilities covering both local communication and communications to the cloud, cloud processing support that includes databases and analytics, user feedback mechanisms, and runtime monitoring to notify researchers as soon as any problems with the system or associated data recording experiences any problems. Having such a platform will also improve the robustness of the system as it is used in many different studies.

---

[7] (2016, Nov 17). Health and Social Conditions of the Poorest Versus Wealthiest ... -NCBI. https://www.ncbi.nlm.nih.gov/pubmed/27854531. Accessed 6 Dec. 2016.

Another issue is that studies for similar issues, e.g., those studying depression, cannot easily share and compare results. Standards for data descriptions, reporting, privacy, and availability are necessary so that researchers can utilize results from prior studies and improve the results in subsequent studies.

**Collaboration Challenges: Data, Expertise, Personnel, and Interventions:** While there is a growing and improving collaboration between technologists and healthcare professionals, paradigm mismatches still exist. Some of the barriers to successful collaboration include disciplinary silos, training of personnel, accessing expertise, availability of data, challenges to initiating new projects.

Enormous amounts of medical data have been collected in the past. Access to that data is often very limited, partly due to how the studies specified policies for their Institutional Review Boards (IRBs). Data formats and use of English sentences for annotations also complicate the ability to reuse the data. With new data mining, machine learning, and natural language processing capabilities, having access to this past data would be invaluable. Policy and data formatting issues need to be resolved to accelerate the pace of discovery and innovation in this field.

In addition, increased training of multidisciplinary personnel and access to appropriate expertise are critical to break down existing disciplinary silos to better facilitate closer interaction between researchers and practitioners of different backgrounds, education, and experiences. Finally, identification of mechanisms to enable collaboration, such as face-to-face interaction, special theme-based workshops, scheduled meetings, etc. is critical to facilitate inter-disciplinary research.

**Healthcare Disparities:** There are acknowledged challenges and tradeoffs in addressing healthcare disparities in smart and pervasive health research programs. For example, research focused on new technology innovation may have limited access and resources to simultaneously engage with diverse population groups. However, it is important for any research to appropriately integrate new information and long-standing knowledge of target populations. The assessment of smart and pervasive health research against intended uses and patient target populations is part of the intellectual merit of the work. Simultaneously, any work that does focus on specific communities or population engagement should adopt and publish appropriate methodologies (e.g., user and population-centered design principles; community-based participatory design; action-based research[8]) on how they effectively engage these groups in the research process to identify their values, assess the potential of technical approaches, and address their concerns in any technology development or other research initiative. Due to the complex and nonlinear nature of working with human populations, robust study designs should also acknowledge how the research process may change if and when different groups and outcomes are included (or discluded) in the process. Additionally, researchers must acknowledge what the limitations are of known missing data regarding target populations.

Research programs should have a commitment to engage with target populations with health disparities from conception of ideas through design, deployment, and evaluation. We encourage funding solicitations to require appropriate methods for design, deployment, and continued engagement with intended target populations as part of the

---

[8] (2011, July 1). The relationship of action research to human ... - ACM Digital Library. Retrieved December 6, 2016, from http://dl.acm.org/citation.cfm?id=1993065

*Intellectual Merit* or *Specific Aims*, being acutely aware of groups who have been and are affected by disparities. It must be kept in mind that disparities in economics, education and other areas can all result in health disparities. We further note that including populations with health disparities contributes to the science itself, because including these communities impacts the generalizability, usability, and appropriation of these systems.

Proposals can specifically address disparities in relation to identifying:
- Access to resources
- Inequities within, among, and between communities
- Needs, designing, building, deploying, evaluating, or meta analysis of various data sets
- Methods to engage people with health disparities (e.g., community-based participatory design,[9] health activism[10]) throughout the research process
- Technical approaches to navigate different resources (e.g., rural communities with limited internet connectivity)

Advocating for including populations at risk for healthcare disparities is necessary but not sufficient. The research community needs (1) resources, (2) expertise, and (3) methods to ensure the contributions generated through this broadening research initiative can improve science through increasing generalizability and thus the impact to science. In terms of resources, the community needs mechanisms and partnership models to sustain engagement throughout the research process and between grant funding mechanisms. Regarding expertise, the computing community needs mechanisms - for researchers at in all career levels - to facilitate connecting with researchers who have valuable knowledge about populations associated with healthcare disparities. Finally, the computing research community needs to combine the needs of resources and expertise to adopt, appropriate, and build methods to design, build, and study sociotechnical systems to address health disparities.

## 7.0      Recommendations

In addition to pursuing the research and development paths described in this paper, we want to call out these specific recommendations for the successful implementation of an impactful smart and pervasive health agenda that results in improved health outcomes, cost-effective healthcare delivery, and improved patient satisfaction, engagement and utilization:

1. New health data platforms are needed to support holistic and unconventional data streams. Current industry platforms under development are either proprietary or implement a limited range of physiological data fields, severely limiting the platforms' utility beyond their immediate application goals. New platforms need to accept data produced from both internal and external sources (including wearables, traditional health monitors, deployable sensors and self-report) with varying levels of noise, fidelity, and sampling rates. These platforms need to support data triage and inspection capabilities to help ensure that data for an individual is being representing appropriately in statistical models. Data quality can easily vary from person to person and device to device.
2. As mobile 24/7 technology is brought to bear on personalized human centric care, new efforts are required that fundamentally address human-in-the-loop participation based on principled psychological, behavioral, and physiological models which are exposed as explicit components of feedback loops. New multidisciplinary research efforts among control theorists, interaction designers, and health professionals are recommended.
3. Rapid progress in infrastructure to support human centric healthcare can only occur if a standardized, composable architecture for integrating devices and data is developed. Further, instantiations of this architecture must support privacy, security, dependability, safety, reusability, and be useable by non-experts. Significant new research efforts in achieving these goals are recommended.
4. Mobile technology offers unprecedented opportunities to improve prevention of chronic diseases. Realizing this promise requires development of new models and theories for just-in-time intervention. In particular, research is needed to better understand intervention timing, content, and delivery modalities and how to optimize their effect to achieve sustained behavior change in the long-term.
5. At the device level, revolutionary approaches in sensing using nano-structures, quantum computing, and non-intrusive `wearable' (e.g. "smart skin") would likely offer next-generation technologies for inexpensive, pervasive, and individualized health monitoring. In addition, research efforts around repurposing non-health sensors for health sensing have started to emerge, such as using the camera/flash combination on smartphones for non-invasive blood screening or the microphone for pulmonary assessment. Leveraging commodity low-cost sensors require the exploration and application of new signal processing and machine learning techniques. Complementary work aimed at increasing sensing capabilities on a smartphone coupled with creative uses of existing sensors will advance smartphones, together with smart sensing, as a compelling platform for personal health management.
6. With the proliferation of healthcare apps and in-home technology, individuals are becoming more involved with their own health decisions, especially in rural communities. Due to many confounding factors, possible simultaneous treatments, and lack of medical expertise of patients there is a great potential for unsafe actions due to conflicts in treatments, including drug-drug interactions. New population-based solutions must be developed to create sound, evidence-based methodologies and to ensure safety in human centric systems.
7. Substantial and sustained collaboration is needed between healthcare practitioners and smart and pervasive health researchers to fully understand the potential of these approaches for new paradigms of healthcare delivery and to identify and address barriers to successful implementation.
8. We encourage the funding agencies to consider health disparities as an integral consideration in their scientific research portfolios and agendas, and to encourage appropriate methods for design, deployment, and continued engagement for the intended target populations, being acutely aware of groups who have been and are affected by disparities.

---

[9] (2015, July 30). Integrating community-based participatory research and informatics .... Retrieved December 6, 2016, from https://www.ncbi.nlm.nih.gov/pubmed/26228766

[10] (2012, May 5). Health Promotion as Activism: Building Community Capacity to Effect .... Retrieved December 6, 2016, from https://pdfs.semanticscholar.org/8617/5302ec9f77df7eedd55c62033dced4efea17.pdf

*This material is based upon work supported by the National Science Foundation under Grant No. 1136993. Any opinions, findings, and conclusions or recommendations expressed in this material are those of the authors and do not necessarily reflect the views of the National Science Foundation.*